\documentclass[reqno]{amsart}
\usepackage{amssymb}
\usepackage{upref}
\begin{document}
\title[On representation of the P--Q pair solution]{On representation of the
P--Q pair solution at the singular point neighborhood}
\author{N V Ustinov}
\thanks{Supported by RFFI grant \symbol{242} 97--01--00752}
\address{Department of Theoretical Physics, Kaliningrad State University,
Al. Nevsky street 14, 236041, Kaliningrad, Russia}
\begin{abstract}
The compatible expansion in series of solutions of both the equations of
P--Q pair at neighborhood of the singular point is obtained in closed form for
regular and irregular singularities.
The conservation laws of the system of ordinary differential equations to
arise from the compatibility condition of the P--Q pair are derived.
\end{abstract}
\maketitle
\section{ Introduction }
The inverse monodromic (or isomonodromic) transformation (IMT) method
\cite{1,2} is a powerful tool for studying the class of nonlinear ordinary
differential equations (ODE's) representable as the compatibility condition of
the overdetermined linear system (P--Q pair).
The IMT method reduces the initial value problem for the system of nonlinear
ODE's to solving the inverse problem for associated isomonodromic linear
equation.
This inverse problem is formulated in terms of the monodromy data, which are
constructed using the asymptotic expansions of solution of the \mbox{P--Q}
pair at neighborhood of the singular points.
It was shown for particular cases that the matrix coefficient of the
isomonodromic equation can be uniquely specified from the monodromy properties
of its global solution \cite{3}.
The existence of the global solution can be investigated in the frameworks of
the theory of the Riemann--Hilbert problem \cite{2}.

This work is devoted to obtaining in closed form of the expansion in
series of solution of P--Q pair at neighborhood of the singular points.
The problem considered here is important for developing the IMT method on
\mbox{P--Q} pairs of arbitrary matrix dimension with different types of
singularities of both the equations forming the pair.
In particular, the compatibility of the equations, that are imposed on the
remainder term of the asymptotic expansion of the \mbox{P--Q} pair solution,
is suggested in implementing the direct problem of the IMT method for the
monodromy data to be determined.
The irregular and regular singularities of the second equation of P--Q pair
(\mbox{Q--equation}) are studied in Sec.III and Sec.IV respectively.
We require no the additional conditions on the coefficient of \mbox{Q--equation}
such as the inequality (on modulo integers for the regular singularity) of the
eigenvalues of the leading coefficients of the expansions in singular points
(cf. \cite{3}).
Besides, the independent variable of the first equation of P--Q pair is not
supposed to be immediately connected with the subset of the "deformation
parameters" of the monodromy data (see Ref.\cite{3}).
The theorems establishing the existence of the compatible expansion in series
at the singular point neighborhood of solutions of the P--Q pair equations are
proven.
The conservation laws for system of nonlinear ODE's admitting the
compatibility condition representation are derived from the expansions in
series of solution of corresponding P--Q pair.

\section{ P--Q pair }
Nonlinear ODE's considered in the frameworks of the IMT method can be written
in the form
\begin{equation}
Q_x - P_\lambda + [Q,P] = 0
\label{1}
\end{equation}
with matrices $P=P(x,\lambda)$ and $Q=Q(x,\lambda)$ depending rationally on
variable $\lambda$.
This equation is the compatibility condition of overdetermined linear system
\begin{equation}
\Psi_x = P \Psi ,
\label{2}
\end{equation}
\begin{equation}
\Psi_\lambda = Q \Psi .
\label{3}
\end{equation}
The six Painlev$\acute{\mbox{\rm e}}$ equations \cite{4} are the most famous
nonlinear ODE's that admit representation (\ref{1}) \cite{3}.

Let at least Q--equation of P--Q pair (\ref{2},\ref{3}) at vicinity of point
$x=x_0$ have the singularity in point \mbox{$\lambda = 0$}.
We also suppose that the coefficients of the P--Q pair are expanded in series
at neighborhood of point $(x=x_0, \lambda=0)$ as given
\begin{equation}
P = \sum_{i=m}^{\infty} \lambda^i P^{(i)} , \,\,\,
Q = \sum_{i=n}^{\infty} \lambda^i Q^{(i)}
\label{4}
\end{equation}
($m \le 0$, $n<0$), where matrix coefficients $P^{(i)}$ and $Q^{(i)}$ are
holomorphic in variable $x$.
Substituting expansions (\ref{4}) into Eq.(\ref{1}) and equalizing to zero the
coefficients at different powers of $\lambda$, one obtains an infinite set of
equations
\begin{equation}
Q_x^{(i)} -(i+1)P^{(i+1)} + \sum_{j=-\infty}^{\infty}
[Q^{(j)},P^{(i-j)}] = 0 \,\,\, (i \ge m+n).
\label{5}
\end{equation}
It is assumed hereafter that $Q^{(i)}=0$ if $i<n$ and $P^{(i)}=0$ if $i<m$.

We quote no system (\ref{2},\ref{3}) as Lax pair to distinguish from the
overdetermined systems for nonlinear partial differential equations integrable
in the frameworks of the inverse scattering (or spectral) transformation
method \cite{5}.
The P--Q pairs, whose coefficients are represented as given
$$
P = \sum_{i=m}^{\infty} (\lambda + f(x))^i P^{(i)}(x) , \,\,\,
Q = \sum_{i=n}^{\infty} (\lambda + f(x))^i Q^{(i)}(x) ,
$$
are evidently led to the form considered here by means of changing independent
variables $(x,\lambda) \to (x, \lambda + f(x))$.

\section{ Irregular singularity of Q--equation }
If point $\lambda=0$ is the irregular singularity of the second equation of
P--Q pair we have the following

\noindent
{\bf Theorem 1.} {\it
Let the asymptotic expansion in series of solution of Eq.(\ref{3}) for fixed
$x=x_0$ at neighborhood of irregular singular point $\lambda=0$ be represented
in closed form
\begin{equation}
\Psi = \sum_{i=0}^{\infty} \lambda^i C^{(i)} \Lambda ,
\label{6}
\end{equation}
where $C^{(0)}=E$, $\Lambda$ is nondegenerate solution of equation
\begin{equation}
\Lambda_{\lambda} = \sum_{i=n}^{-1} \lambda^i \Omega^{(i)} \Lambda
\label{7}
\end{equation}
in point $x=x_0$, coefficients $C^{(i)}$ $(i>0)$ and
$\Omega^{(i)}$ $(n\le i<0)$ are defined from infinite set of equations
\begin{equation}
(i+1) C^{(i+1)} + \sum_{j=n}^{-1} C^{(i-j)} \Omega^{(j)}
- \sum_{j=n}^{\infty} Q^{(j)} C^{(i-j)} = 0 \,\,\,
(i \ge n, C^{(i)}=0 \,\, \mbox{if} \,\, i<0 )
\label{8}
\end{equation}
with $x=x_0$.
Then the compatible asymptotic expansion in series of the solutions of both
the equations of P--Q pair (\ref{2},\ref{3}) at neighborhood of point
$(x=x_0, \lambda=0)$ is represented in form (\ref{6}), where $\Lambda$ is
nondegenerate solution of Eq.(\ref{7}) and equation
\begin{equation}
\Lambda_{x} = \sum_{i=m}^{0} \lambda^i \Phi^{(i)} \Lambda ,
\label{9}
\end{equation}
coefficients $C^{(i)}$ satisfy Eqs.(\ref{8}), which define simultaneously
coefficients $\Omega^{(i)}$, and the set of equations
\begin{equation}
C_x^{(i)} + \sum_{j=m}^{0} C^{(i-j)} \Phi^{(j)}
= \sum_{j=m}^{\infty} P^{(j)} C^{(i-j)} \,\,\,
(i \ge m)
\label{10}
\end{equation}
that also define coefficients $\Phi^{(i)}$ $(m\le i\le 0)$.
There exist matrices $C^{(i)}$, $\Phi^{(i)}$ and $\Omega^{(i)}$, which
solve Eqs.(\ref{8},\ref{10}), such, that the overdetermined system of
equations (\ref{7},\ref{9}) is compatible.
}

\noindent
{\bf Proof.}
From the sets of equations (\ref{8}) and (\ref{10}) for $n \le i <0$ and
$m \le i \le 0$ respectively we have the recurrent definitions of matrices
$\Omega^{(i)}$ and $\Phi^{(i)}$:
\begin{equation}
\Omega^{(i)} = Q^{(i)} + \sum_{j=n}^{i-1} \left( Q^{(j)} C^{(i-j)} -
C^{(i-j)} \Omega^{(j)} \right) ,
\label{11}
\end{equation}
\begin{equation}
\Phi^{(i)} = P^{(i)} + \sum_{j=m}^{i-1} \left( P^{(j)} C^{(i-j)} -
C^{(i-j)} \Phi^{(j)} \right) .
\label{12}
\end{equation}
The normal system of ODE's on coefficients
$C^{(i)}$
\begin{equation}
C_x^{(i)} =
\sum_{j=m}^{i} P^{(j)} C^{(i-j)} - \sum_{j=m}^{0} C^{(i-j)} \Phi^{(j)}
\label{13}
\end{equation}
follows from Eqs.(\ref{10}) if $i>0$.

We assume for convenience that $\Phi^{(i)}=0$ if $i<m$ or $i>0$,
$\Omega^{(i)}=0$ if $i<n$ or $i\ge0$.
This agreement yields immediately two sets of useful identities:
\begin{equation}
\Omega^{(i)} = Q^{(i)} + \sum_{j=-\infty}^{i-1} \left( Q^{(j)} C^{(i-j)} -
C^{(i-j)} \Omega^{(j)} \right) \,\,\, (i<0),
\label{14}
\end{equation}
\begin{equation}
\Phi^{(i)} = P^{(i)} + \sum_{j=-\infty}^{i-1} \left( P^{(j)} C^{(i-j)} -
C^{(i-j)} \Phi^{(j)} \right) \,\,\, (i\le0)
\label{15}
\end{equation}
from Eqs.(\ref{11},\ref{12}).
Using notations
\begin{equation}
H^{(i)} = \Omega_x^{(i)} -(i+1)\Phi^{(i+1)} + \sum_{j=-\infty}^{-1}
[\Omega^{(j)},\Phi^{(i-j)}] \,\,\, (i<0) ,
\label{16}
\end{equation}
the system of equations arising from the compatibility condition of
Eq.(\ref{7}) and Eq.(\ref{9}) is written in following manner:
\begin{equation}
H^{(i)} = 0 \,\,\, (m+n \le i < 0) .
\label{17}
\end{equation}
This equation is obviously valid for $i<m+n$.

Let matrices $C^{(i)}$ $(i>0)$ satisfy system (\ref{13}) at neighborhood of
point \mbox{$x=x_0$}.
Substitution of identities (\ref{14},\ref{15}) into first two terms in the
right--hand side of Eqs.(\ref{16}) leads after cumbersome calculations to
formulas:
\begin{equation}
H^{(i)} = \sum_{j=-\infty}^{i} \left( F^{(i-j)} \Phi^{(j)} - P^{(j)} F^{(i-j)}
\right) - \sum_{j=-\infty}^{i-1} C^{(i-j)} H^{(j)} ,
\label{18}
\end{equation}
where
\begin{equation}
F^{(i)} = (i+1) C^{(i+1)} + \sum_{j=-\infty}^{-1} C^{(i-j)} \Omega^{(j)}
- \sum_{j=-\infty}^{i} Q^{(j)} C^{(i-j)} \,\,\, (i \ge 0).
\label{19}
\end{equation}
It should be stressed that the expression for $F^{(i)}$ is nothing but the
left--hand side of Eq.(\ref{8}) for $i \ge 0$.
Differentiation of Eqs.(\ref{19}) gives, taking into account
Eqs.(\ref{5},\ref{13},\ref{16}), system of ODE's:
\begin{equation}
F_x^{(i)} = \sum_{j=-\infty}^{i} \left( P^{(j)} F^{(i-j)} -
F^{(i-j)} \Phi^{(j)} \right) + \sum_{j=-\infty}^{-1} C^{(i-j)} H^{(j)} .
\label{20}
\end{equation}

Since Eqs.(\ref{8}) are valid for fixed $x=x_0$, we can choose the initial
values of matrices $C^{(i)}$ $(i>0)$ in this point such, that
\begin{equation}
F^{(i)} = 0
\label{21}
\end{equation}
for $x=x_0$ (see the remark after Eqs.(\ref{19})).
Then Eqs.(\ref{21}) are fulfilled at a neighborhood of point $x=x_0$ due to
Eqs.(\ref{18},\ref{20}).
Vanishing right--hand side of Eqs.(\ref{18}) provides for the compatibility of
overdetermined linear system (\ref{7},\ref{9}), whose coefficients are defined
by Eqs.(\ref{11},\ref{12}).
We have next from Eqs.(\ref{11},\ref{19},\ref{21}) that matrices $C^{(i)}$
and $\Omega^{(i)}$ satisfy Eqs.(\ref{8}).
At last, it is checked by direct substitution that expansion (\ref{6}) yields
formally the solution of P--Q pair.
\rule{5pt}{5pt}

If $m=0$ the proof of theorem can be carried out in more simple way by
supposing that matrices $\Omega^{(i)}$ are solutions of Eqs.(\ref{17}).
Eqs.(\ref{13}) can be solved recurrently in this case.
Choosing initial values for $C^{(i)}$ and $\Omega^{(i)}$ to satisfy
Eqs.(\ref{8}), one deduces from Eqs.(\ref{16},\ref{17}) the matrix
conservation laws of the system of nonlinear ODE's, which admits
representation (\ref{1}).
By means of nondegenerate matrix solution $\Psi_0$ of linear equation
\begin{equation}
\Psi_{0,\,x} = P^{(0)} \Psi_0
\label{22}
\end{equation}
conservation laws $J^{(i)}$ $(n \le i < 0)$ are written in next form:
$$
J^{(i)} = \Psi_0^{-1} \Omega^{(i)} \Psi_0 .
$$

For $m<0$ the matrix conservation law of corresponding system of nonlinear
ODE's will be given by matrix $J^{(-1)}$ if $\Psi_0$ is solution of
Eq.(\ref{22}) with coefficient $\Phi^{(0)}$ instead of $P^{(0)}$.

Conservation laws $J^{(-1)}$ and $\Psi_0^{-1} Q^{(-1)} \Psi_0$, which
corresponds to the regular singularity discussed in the sequel, are connected
with "exponent matrix of formal monodromy" (see Ref.\cite{3})
if the eigenvalues of matrix $Q^{(n)}$ are distinct and $n<m\le0$.
Explicit expression for matrix $\Lambda$ can be then obtained and independent
variable $x$ can be directly associated with the subset of the deformation
parameters of the monodromy data \cite{3}.

\section{ Regular singularity of Q--equation }
In this section we consider the case $n=-1$.
Expansion in series of solution of Eq.(\ref{3}) for fixed $x=x_0$ at
neighborhood of singular point $\lambda=0$ is represented in closed form:
\begin{equation}
\Psi = \sum_{i=0}^{\infty} \sum_{j=0}^{N-1} \lambda^i (\ln \lambda)^j
C^{(i,j)} \Lambda .
\label{23}
\end{equation}
Here $N$ is the matrix dimension of P--Q pair, $C^{(0,0)}=E$,
$\Lambda$ is nondegenerate solution of equation
\begin{equation}
\Lambda_{\lambda} = \lambda^{-1} Q^{(-1)} \Lambda
\label{24}
\end{equation}
in point $x=x_0$, coefficients $C^{(i,j)}$
($i \ge 0$, $0 \le j < N$, $i^2 + j^2 \ne 0$) are defined from infinite set of
equations
\begin{equation}
(i+1) C^{(i+1,j)} + [C^{(i+1,j)}, Q^{(-1)}] + (j+1) C^{(i+1,j+1)} -
\sum_{k=0}^{\infty} Q^{(k)} C^{(i-k,j)} = 0
\label{25}
\end{equation}
($i \ge -1$, $0 \le j < N$, $C^{(i,j)}=0$ if $i<0$ or $j \ge N$) with $x=x_0$.

The theorem of the previous section is valid if the expansion in series for
fixed $x=x_0$ at neighborhood of regular singular point $\lambda=0$ has form
(\ref{6}).
It means that conditions $C^{(i,j)}=0$ ($i \ge 0 , j > 0$) will be kept under
the evolution of coefficients $P^{(i)}$ and $Q^{(i)}$ governed by
compatibility condition (\ref{5}).
The logarithmic terms can enter the expansion in this case through matrix
$\Lambda$.
The proof of analogous theorem for expansion (\ref{23}) containing explicitly
the logarithmic terms encounters difficulties since the expansion possesses
the internal degrees of freedom.
Nevertheless, in the case $m=0$ we come to

\noindent
{\bf Theorem 2.} {\it
The compatible expansion in series of the solutions of both the equations of
P--Q pair (\ref{2},\ref{3}) at neighborhood of point $(x=x_0, \lambda=0)$ is
represented by Eq.(\ref{23}), in which $\Lambda$ is nondegenerate solution of
Eq.(\ref{24}) and equation
\begin{equation}
\Lambda_{x} = P^{(0)} \Lambda ,
\label{26}
\end{equation}
coefficients $C^{(i,j)}$ satisfy Eqs.(\ref{25}) and the set
of equations
\begin{equation}
C_x^{(i,j)} + [C^{(i,j)}, P^{(0)}]
= \sum_{k=1}^{\infty} P^{(k)} C^{(i-k,j)} \,\,\, (i \ge 0, 0 \le j < N) .
\label{27}
\end{equation}
}

\noindent
{\bf Proof.}
It is seen from Eqs.(\ref{5}) that overdetermined system of equations
(\ref{24}) and (\ref{26}) is compatible.
If $\Psi_0$ is nondegenerate solution of Eq.(\ref{22}), then matrix
$\Psi_0^{-1} Q^{(-1)} \Psi_0$ is the conservation law of corresponding system
of ODE's arising from the compatibility condition of P--Q pair
(\ref{2},\ref{3}).

Substitution of expansion (\ref{23}) into P--Q pair gives two sets of
equations (\ref{25},\ref{27}).
Let matrices $C^{(i,j)}$ satisfy system (\ref{27}) at neighborhood of point
\mbox{$x=x_0$}.
Introducing notation for the expression in left--hand side of Eqs.(\ref{25})
$$
F^{(i,j)} = (i+1) C^{(i+1,j)} + [C^{(i+1,j)}, Q^{(-1)}] + (j+1) C^{(i+1,j+1)} -
\sum_{k=0}^{\infty} Q^{(k)} C^{(i-k,j)}
$$
($i \ge -1$, $0 \le j < N$), we obtain
$$
F_x^{(i,j)} + [F^{(i,j)}, P^{(0)}] =
\sum_{k=1}^{i+1} P^{(k)} F^{(i-k,j)} ,
$$
taking into account Eqs.(\ref{5},\ref{27}).
So, the initial values of matrices $C^{(i,j)}$ in point $x=x_0$ can be chosen
to satisfy condition
$$
F^{(i,j)} = 0
$$
at a neighborhood of this point.
\rule{5pt}{5pt}

The problem remaining open is the representation of the compatible expansion
in series in closed form of the solutions of both the P--Q pair equations
if $m<0$ and some coefficients $C^{(i,j)}$ ($j>0$) are nonequal to zero.


\begin{thebibliography}{99}
\bibitem{1} Flaschka H and Newell A C 1980 {\it Commun. Math. Phys.} {\bf 76}
65; 1980 Lecture Notes in Pure and Applied Mathematics {\bf 54} 373
\bibitem{2} Fokas A S and Ablowitz M J 1983 {\it Commun. Math. Phys.}
{\bf 91} 381; Fokas~A~S, Mugan~U and Ablowitz~M~J 1988 {\it Physica} {\bf D30}
247; Fokas~A~S and Zhou~X 1992 {\it Commun. Math. Phys.} {\bf 144} 601;
Fokas~A~S, Mugan~U and Zhou~X 1992 {\it Inverse Problems} {\bf 8} 757
\bibitem{3} Jimbo~M, Miwa~T and Ueno~K 1981 {\it Physica} {\bf D2} 306;
Jimbo~M and Miwa~T 1981 {\it Physica} {\bf D2} 407; {\bf D4} 26
\bibitem{4} Painlev$\acute{\mbox{\rm e}}$ P 1900 {\it Bull. Soc. Math. Fr.}
{\bf 28} 214; 1902 {\it Acta Math.} {\bf 25} 1;
Gambier~B 1909 {\it Acta Math.} {\bf 33} 1
\bibitem{5} Lax P D 1968 {\it Commun. Pure and Appl. Math.} {\bf 21} 467
\end{thebibliography}
\end{document}